\documentclass{PoS}
\usepackage{amsmath}

\title{Symmetrizing the signal distribution of radio emission from inclined air showers}

\ShortTitle{Symmetrizing the signal distribution of radio emission from inclined air showers}

\author{\speaker{Tim Huege}$^{ab}$, Felix Schlueter$^a$ and Lukas Brenk$^a$\\
\llap{$^a$} Institut f\"ur Kernphysik, Karlsruhe Institute of Technology (KIT), PO Box 3640, 76021 Karlsruhe, Germany\\
\llap{$^b$} Astrophysical Institute, Vrije Universiteit Brussel, Pleinlaan 2, 1050 Brussels, Belgium\\
E-mail: \email{tim.huege@kit.edu}, \email{felix.schlueter@kit.edu}}

\abstract{Radio detection of inclined air showers currently receives special attention. It can be performed with very sparse antenna arrays and yields a pure measurement of the electromagnetic air-shower component, thus delivering information that is highly complementary to the measurement of the muonic component using particle detectors. However, radio-based reconstruction of inclined air showers is challenging in light of asymmetries induced in the radio-signal distribution by early-late effects as well as the superposition of geomagnetic and charge-excess radiation. We present a model for the signal distribution of radio emission from inclined air showers which allows explicit compensation of these asymmetries. In a first step, geometrical early-late asymmetries are removed. Secondly, a universal parameterization of the charge-excess fraction as a function of the air-shower geometry, the atmospheric density profile and the lateral distance from the shower axis is used to compensate for the charge-excess contribution to the signal. The resulting signal distribution of the pure geomagnetic emission is then fit with a rotationally symmetric lateral distribution function, the area integration of which yields the radiation energy as an estimator for the cosmic-ray energy. We present the details and performance of our model, which lays the foundation for robust and precise reconstruction of inclined air showers from radio measurements.}

\FullConference{36th International Cosmic Ray Conference -ICRC2019-\\
		July 24th - August 1st, 2019\\
		Madison, WI, U.S.A.}

\begin{document}

\section{Introduction}

The signal distribution of radio emission from extensive air showers exhibits pronounced asymmetries arising from the superposition of geomagnetic and charge-excess emission \cite{HuegePLREP}. In an event reconstruction, these asymmetries have to be taken into account either by using a two-dimensional signal distribution \cite{2dLDF}, by treating geomagnetic and charge-excess emission individually \cite{GeoCE}, or by correcting for the asymmetries \cite{TunkaRex}. In inclined air showers, for which reconstruction procedures are still under development, additional asymmetries due to early-late effects \cite{arena} and refractive index effects \cite{GottowikICRC2019} further complicate the situation.  In the following, we present an approach specifically developed for inclined air showers with zenith angles above 60$^{\circ}$ based on symmetrizing the signal distribution and fitting it with a rotationally symmetric function. Our goal is to achieve good reconstruction quality with a function containing only a limited number of parameters.
To develop and evaluate our model, we use a set of 3111 CoREAS\footnote{The version of CoREAS used for this study has been optimized for more precise and efficient calculation of radio emission from extensive air showers.} \cite{CoREAS} simulations for proton and iron primaries, spanning energies from $10^{18.4}$ to $10^{20.2}$\,eV, zenith angles from $65.0^\circ$ to $80.0^\circ$ and 8 equidistantly spaced azimuth angles. The magnetic field and atmosphere as well as observer altitude are fixed to values applicable for the site of the Pierre Auger Observatory. Simulations with geomagnetic angles below 20$^\circ$ have been excluded in this analysis.

\section{Geometrical early-late correction}

\begin{figure}
    \center
    \includegraphics[width=.75\textwidth]{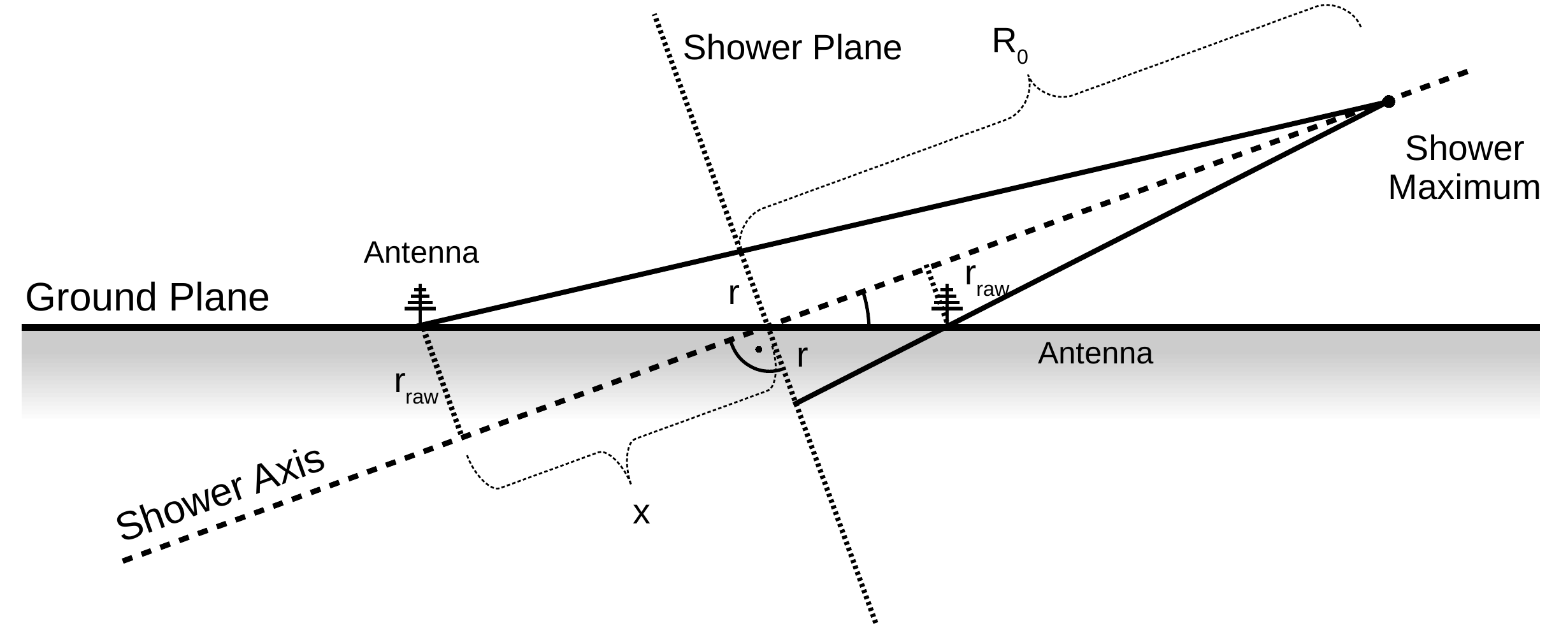}
    \caption{Illustration of the early-late correction geometry. See text for details. From \cite{arena}.}
    \label{fig:early-late-diagram}
\end{figure}

In reference \cite{arena}, we presented a correction for asymmetries arising from the fact that emission above the shower axis propagates on longer lines of sight to the ground than emission below the shower axis. The correction consists of projecting antenna positions to the shower plane along the {\em line of sight from the antenna to a presumed point source at the shower maximum}, as illustrated in figure \ref{fig:early-late-diagram}. This influences the axis distances $r_\text{raw}$ of the antennas, effectively translating them to off-axis angles. Furthermore, the energy fluences $f_\text{raw}$ is corrected with an inverse square dependence on the geometrical distance to a point source at the shower maximum. In summary:
\begin{align}
f = f_\text{raw} \cdot \left(\frac{R}{R_0}\right)^2 \quad \quad &
r = r_\text{raw} \cdot \frac{R_0}{R} \quad \quad R \equiv R_0 + x
\end{align}

\begin{figure}[t]
    \center
    \includegraphics[width=.7\textwidth]{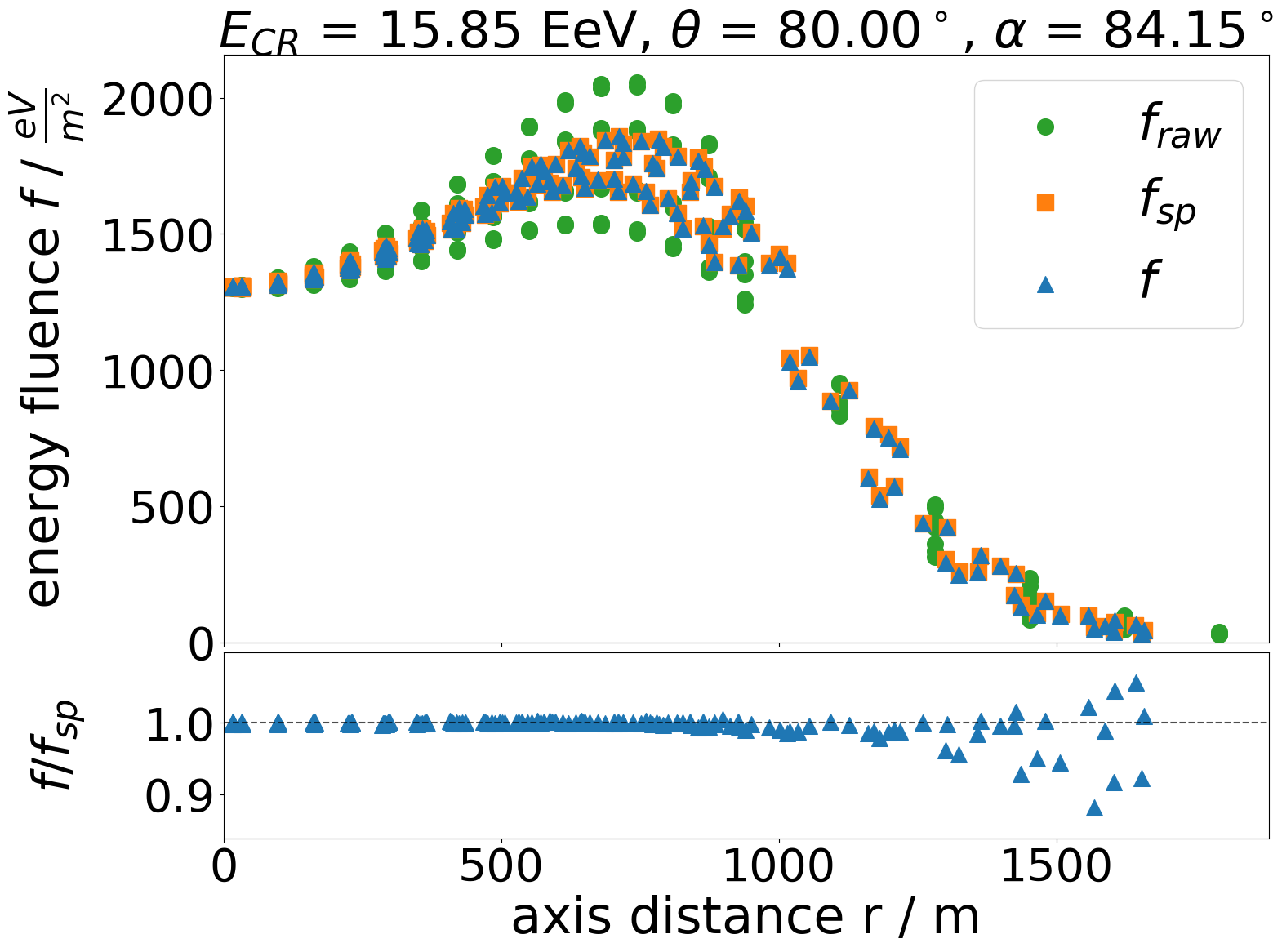}
    \caption{Energy fluence of an air shower simulated with CoREAS in the shower plane ($f_{\mathrm{sp}}$, orange) and the ground plane ($f_{\mathrm{raw}}$, green) as well as after early-late correction ($f$, blue).}
    \label{fig:early-late}
\end{figure}

\begin{figure}[t]
    \center
    \includegraphics[width=.48\textwidth]{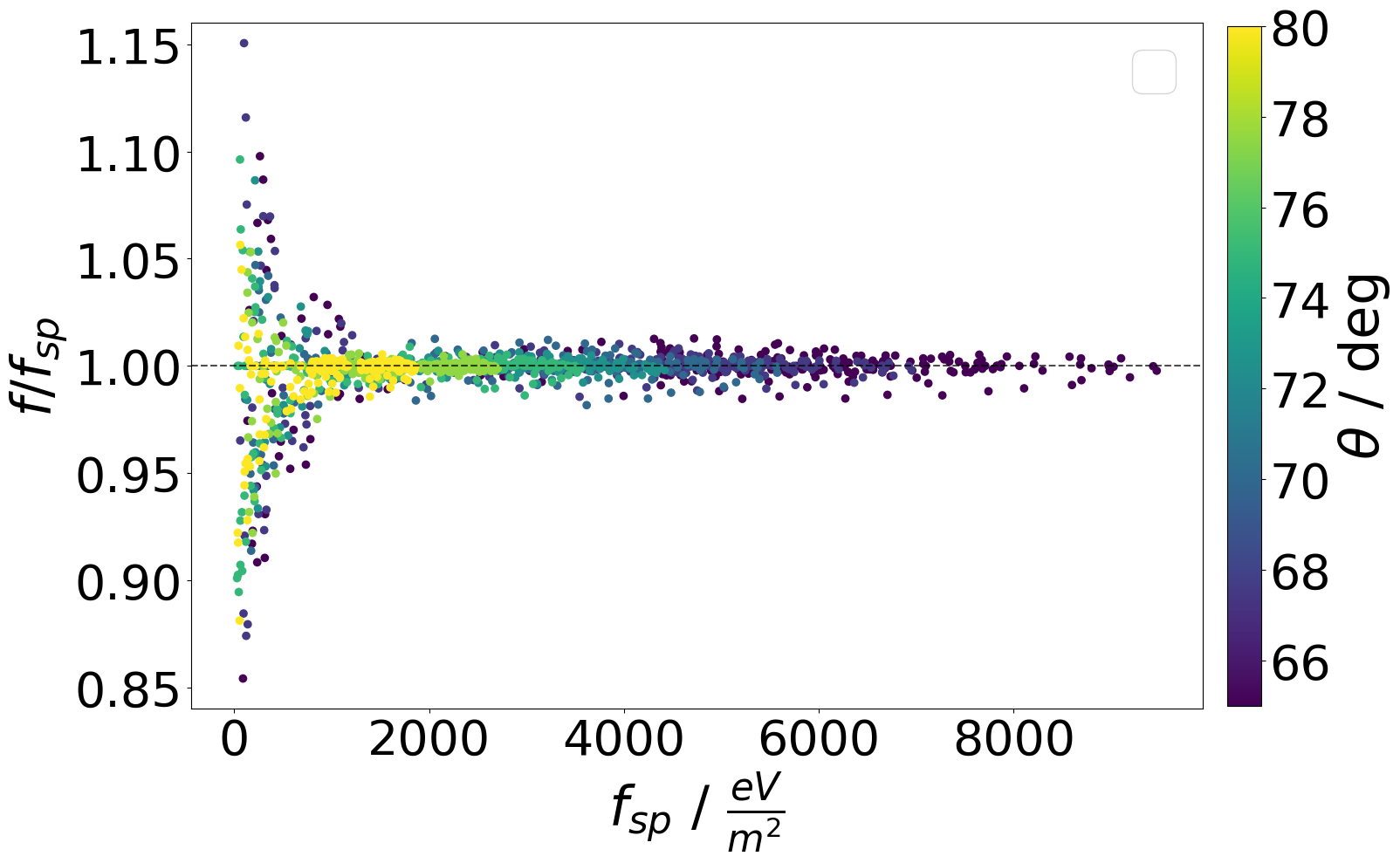}
    \includegraphics[width=.48\textwidth]{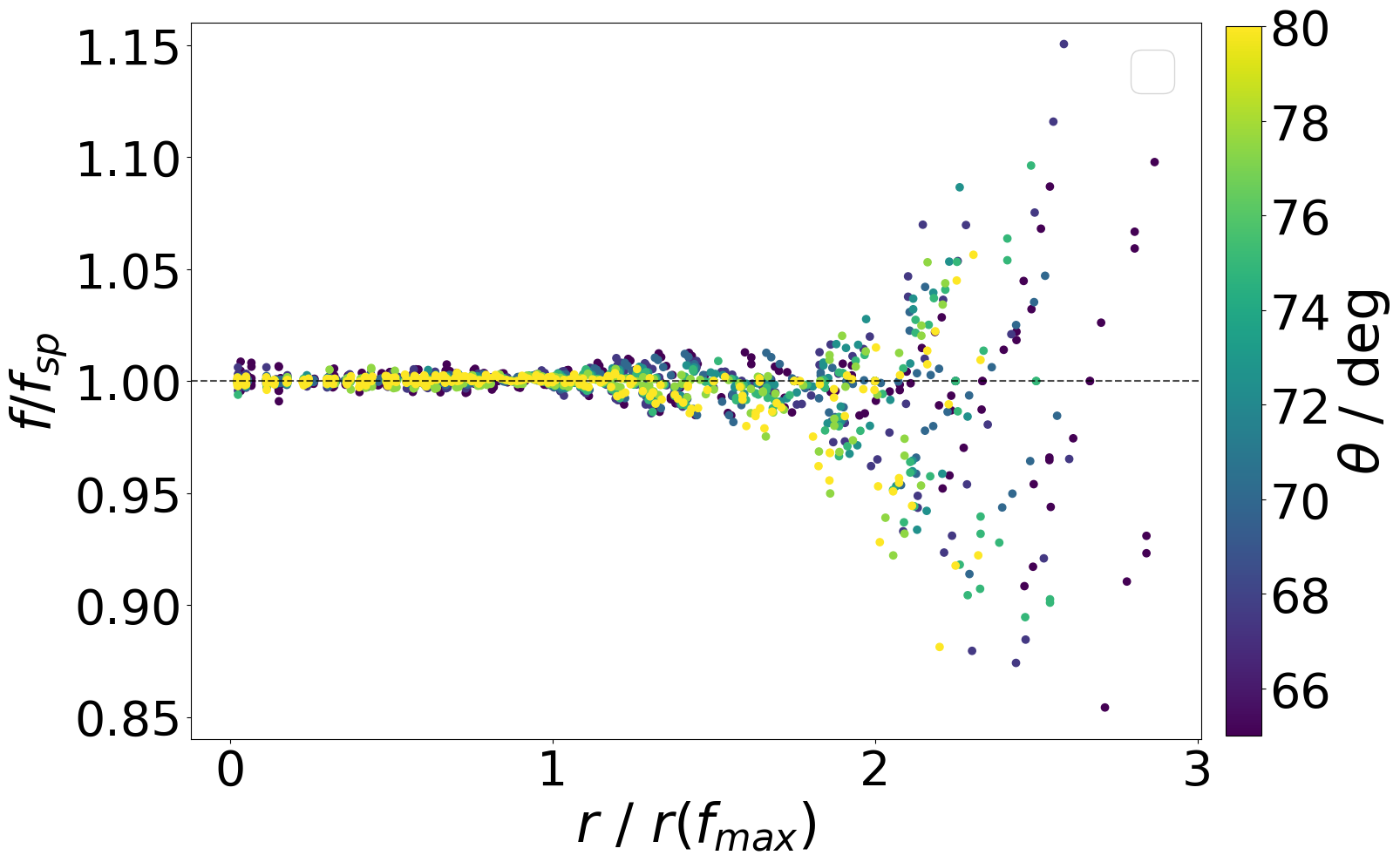}
    \caption{Comparison of the energy fluences $f_{\mathrm{sp}}$ simulated directly in the shower plane and energy fluences $f$ simulated in the ground plane followed by early-late correction to the shower plane. Left: As a function of energy fluence. Right: As a function of axis distance normalized to the axis distance with maximum fluence.}
    \label{fig:early-late2}
\end{figure}

Here, we evaluate the quality of the correction by simulating radio emission with CoREAS both at positions on a star-shape grid in the shower plane and the corresponding ground-plane positions, and then applying the early-late correction to the ground-plane signals using the true depth of the shower maximum. As illustrated in figure \ref{fig:early-late} for one particular geometry, the correction on the ground-plane simulation indeed reproduces the simulation in the shower plane to within 2\% for most axis distances. A comparison for simulations with zenith angles from 65$^\circ$ to 80$^\circ$ is shown in figure \ref{fig:early-late2}. The correction degrades at small values of the energy fluence, which correspond to large axis distances. The degradation 
might be related to additional asymmetries related to refractive index effects \cite{GottowikICRC2019}. Such asymmetries would affect signals in the ground plane differently than signals in the shower plane.

\section{Parameterization of the charge-excess emission}
\begin{figure}[t]
    \center
    \includegraphics[width=.7\textwidth]{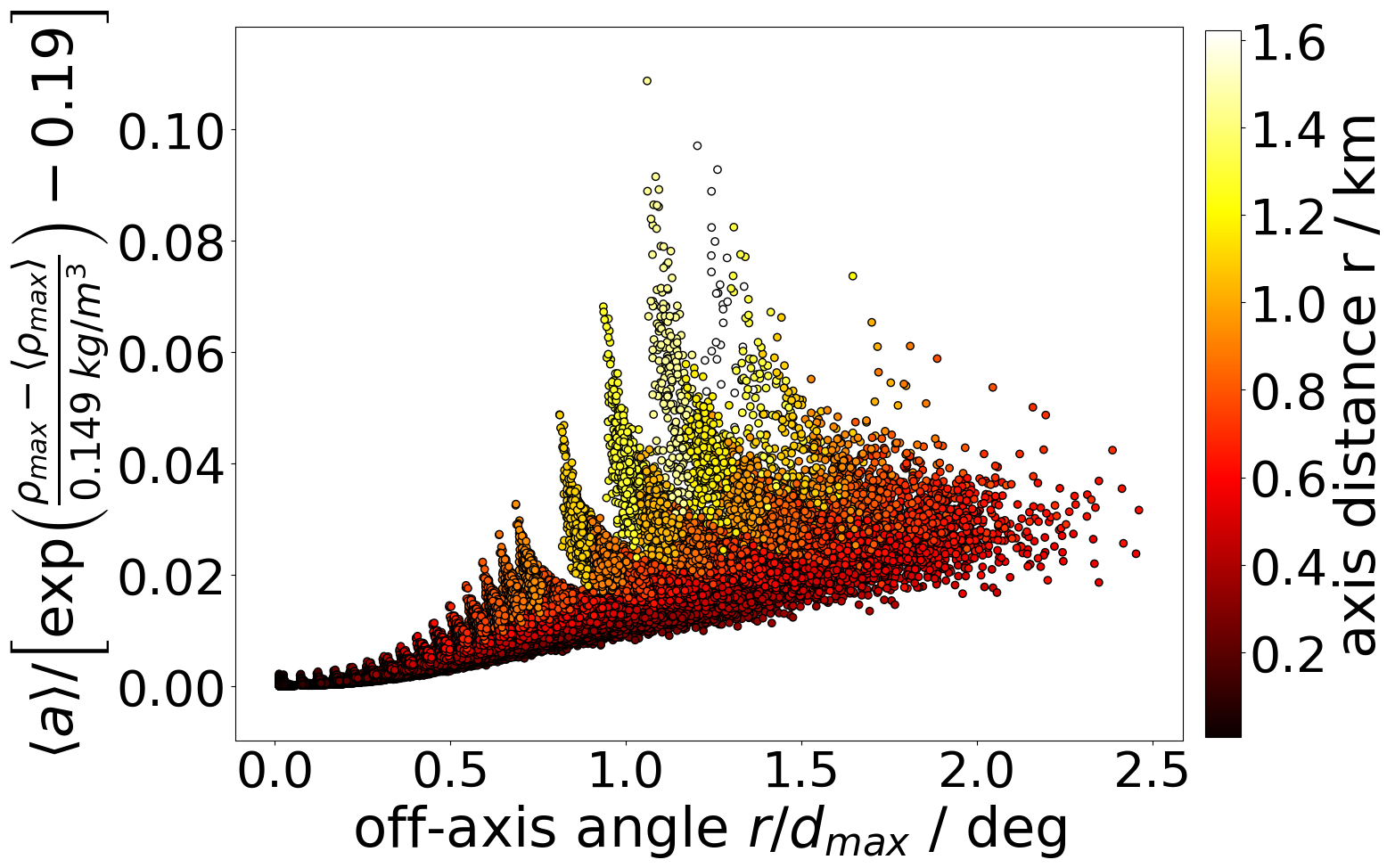}
    \caption{Charge-excess fraction as defined in equation (\ref{eq:ce_fluence}) as a function of the off-axis angle and the antenna axis distance. For each simulated shower, the presented data are averaged over the positions simulated on a concentric ring.}
    \label{fig:charge_excess_fraction}
\end{figure}
Once early-late asymmetries are corrected for, we can remove the asymmetries introduced by the sub-dominant charge-excess contribution interfering with the geomagnetic emission. To do so, we develop a parameterization of the charge-excess fraction 
\begin{equation}
\label{eq:ce_fluence}
    a \equiv \sin^2(\alpha) \cdot \frac{f_{\mathrm{ce}}}{f_{\mathrm{geo}}},
\end{equation}
where $\alpha$ denotes the geomagnetic angle and $f_x$ denotes the energy fluences of the two contributions.\footnote{Please note that this definition of the charge-excess fraction constitutes the square of similar definitions based on ratios of electric field amplitudes.} For each individual simulated position except those on the {\bf v}$\times${\bf B} axis we determine the contributions of the two emission mechanisms exploiting their known polarisation characteristics
\begin{gather}
\label{eq:ce_fluence_pos2}
    f_{\mathrm{geo}}^{\mathrm{pos}} = \left(\sqrt{f_{\textbf{v}\times\textbf{B}}} -  \frac{\cos(\phi)}{|\sin(\phi)|} \cdot \sqrt{f_{\textbf{v}\times\textbf{v}\times\textbf{B}}} \right)^2 \nonumber \\
    f_{\mathrm{ce}}^{\mathrm{pos}} = \frac{1}{\sin^2(\phi)} \cdot  f_{\textbf{v}\times\textbf{v}\times\textbf{B}}. \label{eq:posfraction}
\end{gather}
with {\bf v} the shower axis in direction of the shower particles, {\bf B} the direction of the earth magnetic field, and $\phi$ the polar angle between the {\bf v}$\times${\bf B} axis and observer position \cite{GeoCE}. From the two contributions, we calculate the charge-excess fractions and show them averaged over simulated positions on the same concentric ring in figure \ref{fig:charge_excess_fraction}. We note that small asymmetries within the concentric rings exist which could be parameterized explicitly in the future. Fitting the data of all individual simulated positions in one go, we then find the following parameterization:
\begin{equation}
    a(r, d_{\mathrm{max}}, \rho_{\mathrm{max}}) = 0.373 \cdot \frac{r}{d_{\mathrm{max}}} \cdot \exp\left(\frac{r}{762.6 \; \text{m}} \right) \cdot \left[\exp\left(\frac{\rho_{\mathrm{max}} - \langle \rho_{\mathrm{max}} \rangle}{0.149 \; \text{kg}/\text{m}^3}\right) - 0.189 \right], \label{eq:aparam}
\end{equation}
with (early-late corrected) axis distance $r$, geometrical distance from shower core to shower maximum $d_{\mathrm{max}}$, air density at the shower maximum $\rho_{\mathrm{max}}$ and $\langle \rho_{\mathrm{max}} \rangle = 0.4$\,kg/m$^3$. In comparison with reference \cite{arena}, we now use the geometrical distance to shower maximum instead of the depth of shower maximum, and we no longer include a zenith-angle dependent term but rather correct for the density at shower maximum as originally introduced in reference \cite{GlaserJCAP}. The motivation for these changes is to make the parameterization more universal with respect to the atmospheric model and variations of the observer altitude.
\section{Symmetrizing the signal distribution}
\begin{figure}[t]
    \center
    \includegraphics[width=.7\textwidth]{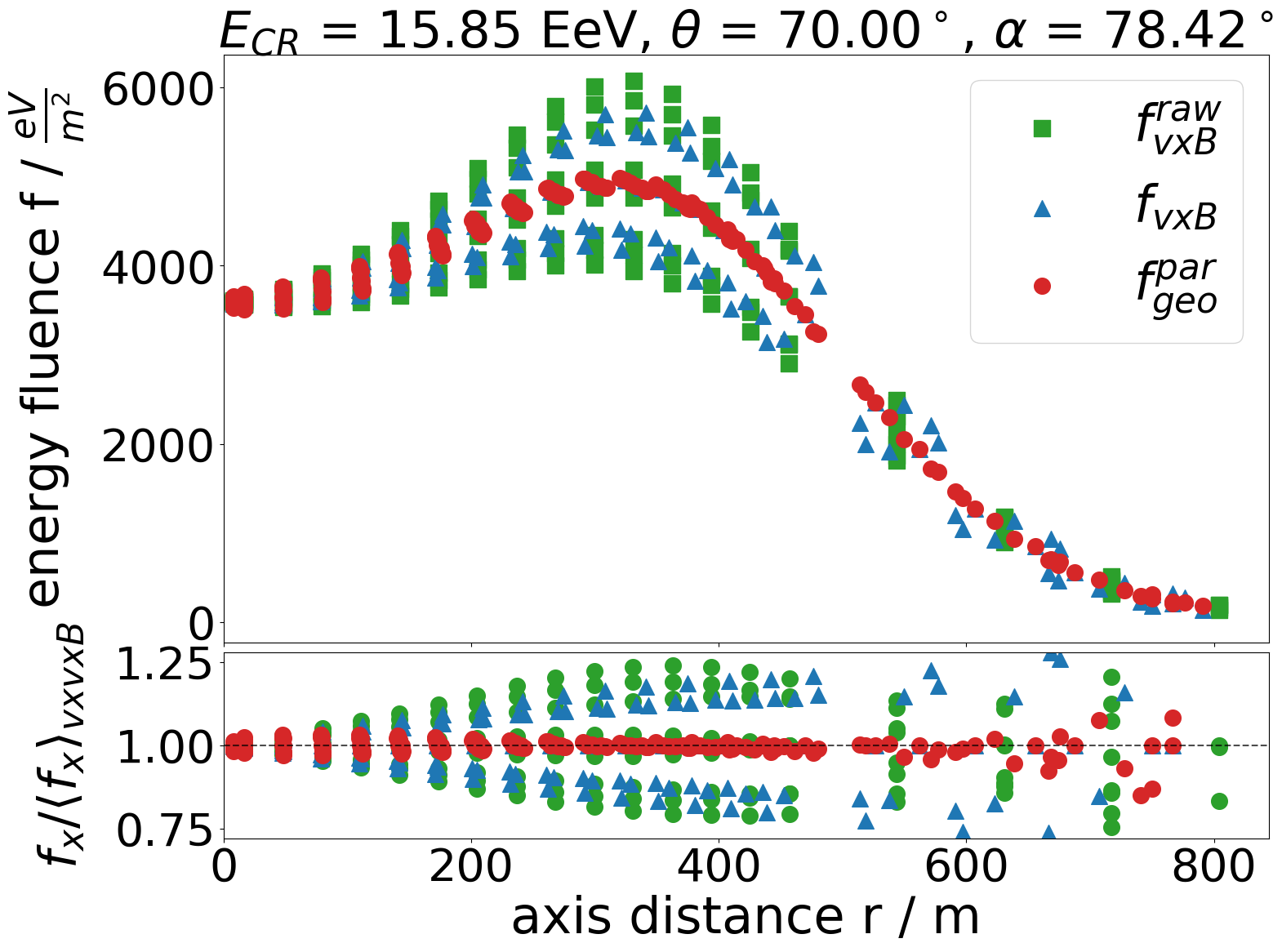}
    \caption{Illustration of the symmetrization of the signal distribution. Uncorrected data in green, early-late corrected data in blue and geomagnetic energy fluence determined with equations (\ref{eq:geomag}) and (\ref{eq:aparam}) in red.}
    \label{fig:symmetrization}
\end{figure}
With our model for early-late correction and the parameterization of the charge-excess fraction, we can now symmetrize the signal distribution of a given air shower by calculating the pure geomagnetic energy fluence. This exploits the known signal polarizations of the two contributions to recover the energy fluence of the pure geomagnetic emission $f_{\mathrm{geo}}$ from a measurement of the signals in the {\bf v}$\times${\bf B} signal polarization according to
\begin{equation}
    f_{\mathrm{geo}}^{\mathrm{par}} = \frac{f_{\textbf{v}\times\textbf{B}}}{\left(1 + \frac{\cos(\phi)}{|\sin(\alpha)|} \cdot \sqrt{a(r, d_{\mathrm{max}}, \rho_{\mathrm{max}})}\right)^2}, \label{eq:geomag}
\end{equation}
where $\phi$ denotes the polar angle of a given antenna position in shower plane coordinates. We note that if sufficient signal is measurable in the {\bf v}$\times${\bf v}$\times${\bf B} polarization, the geomagnetic energy fluence can be determined directly from the measurement according to equation (\ref{eq:posfraction}), without reverting to the parameterization. In figure \ref{fig:symmetrization}, the symmetrization is illustrated. The raw signal distribution on the ground (green squares) is strongly asymmetric with a scatter around the mean of up to $\sim 25\,\%$. After applying the early-late correction (blue triangles), the signal distribution becomes more symmetric with a remaining scatter of $\sim 15\,\%$. Finally, calculating the geomagnetic energy fluence according to equations (\ref{eq:geomag}) and (\ref{eq:aparam}) (red points) results in a mostly rotationally symmetric lateral distribution. The remaining scatter for the example event (figure \ref{fig:symmetrization}) constitutes to $\sim 3\,\%$ close to the shower axis, $\sim 1\,\%$ at the cherenkov angle, and $\lesssim 10\,\%$ far away from the shower axis. We note that the symmetrization is not fully successful in the inner region of the signal distribution, where the parameterization tends to overestimate the charge-excess fraction.

\begin{figure}[t]
    \center
    \includegraphics[width=.48\textwidth]{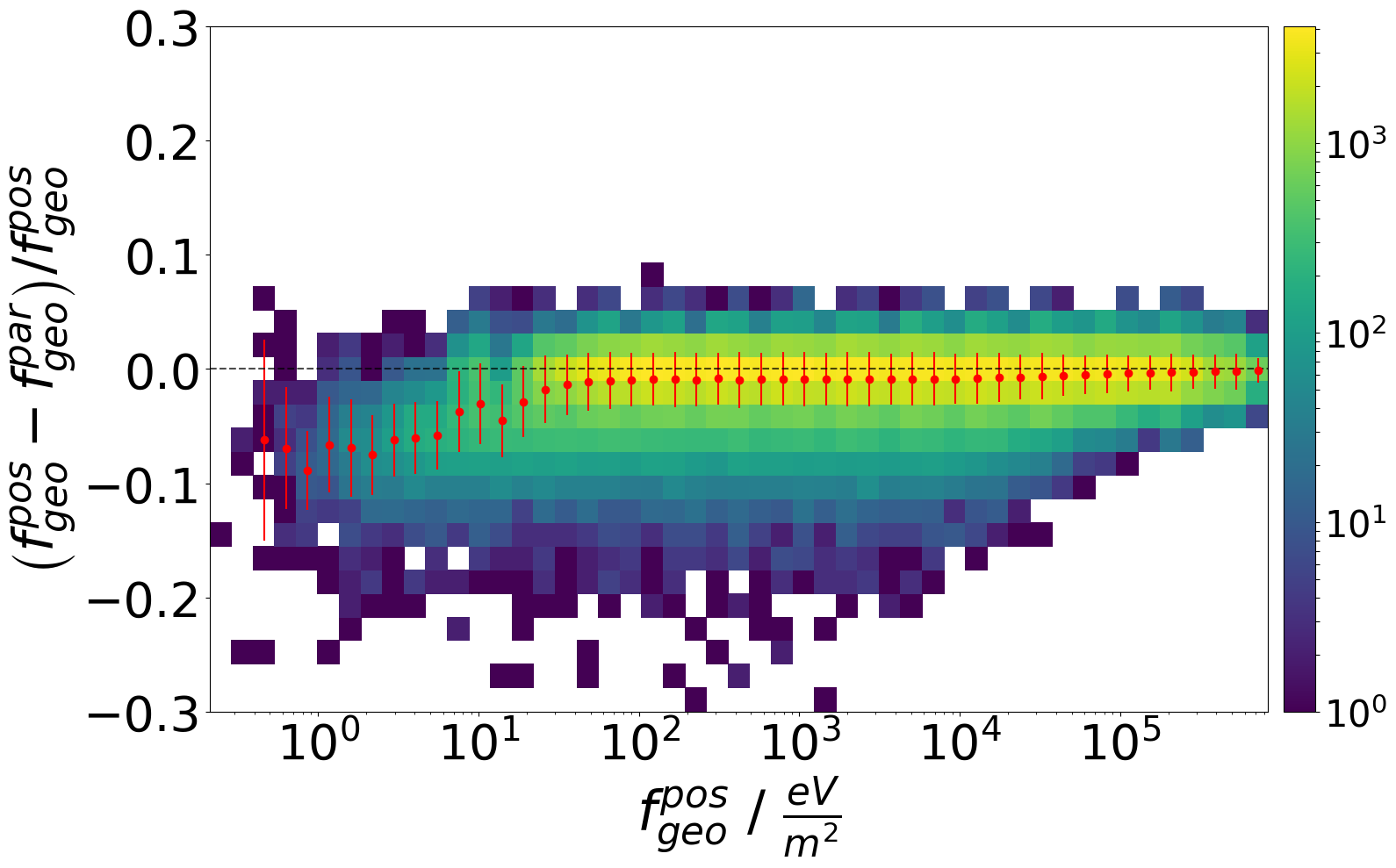}\hfill
    \includegraphics[width=.5\textwidth]{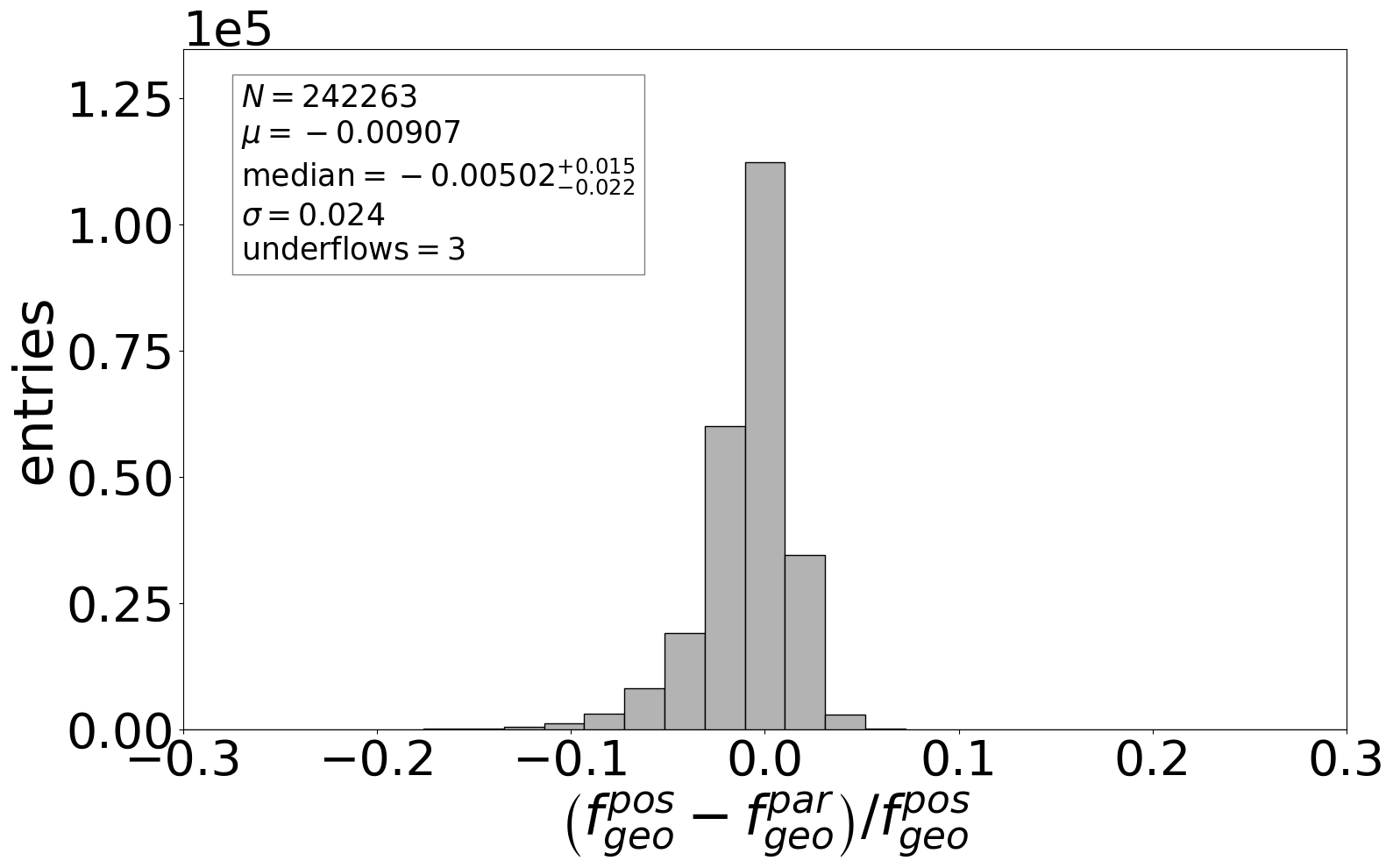}
    \caption{Comparison of the geomagnetic energy fluence $f_{\mathrm{geo}}^{\mathrm{par}}$ determined with the parameterization and the geomagnetic energy fluence $f_{\mathrm{geo}}^{\mathrm{pos}}$ calculated from the signal polarization at each simulated position.}
    \label{fig:ce_valid}
\end{figure}

The quality of the symmetrization is demonstrated more quantitatively in figure \ref{fig:ce_valid}, where we illustrate how the geomagnetic energy fluence as determined using the parameterization via equations (\ref{eq:aparam}) and (\ref{eq:geomag}) compares to the true geomagnetic energy fluence determined at each given position using equation (\ref{eq:posfraction}). An underestimation is visible at low energy fluences, corresponding to signals at large lateral distances. This region of parameter space, however, only contributes minimally in the integration of the energy fluences to determine the radiation energy. A small bias of the order of 1\% is visible also at high fluences, which we consider well within acceptable limits.
\section{Fitting the lateral signal distribution}
\begin{figure}[t]
    \center
    \includegraphics[width=.65\textwidth]{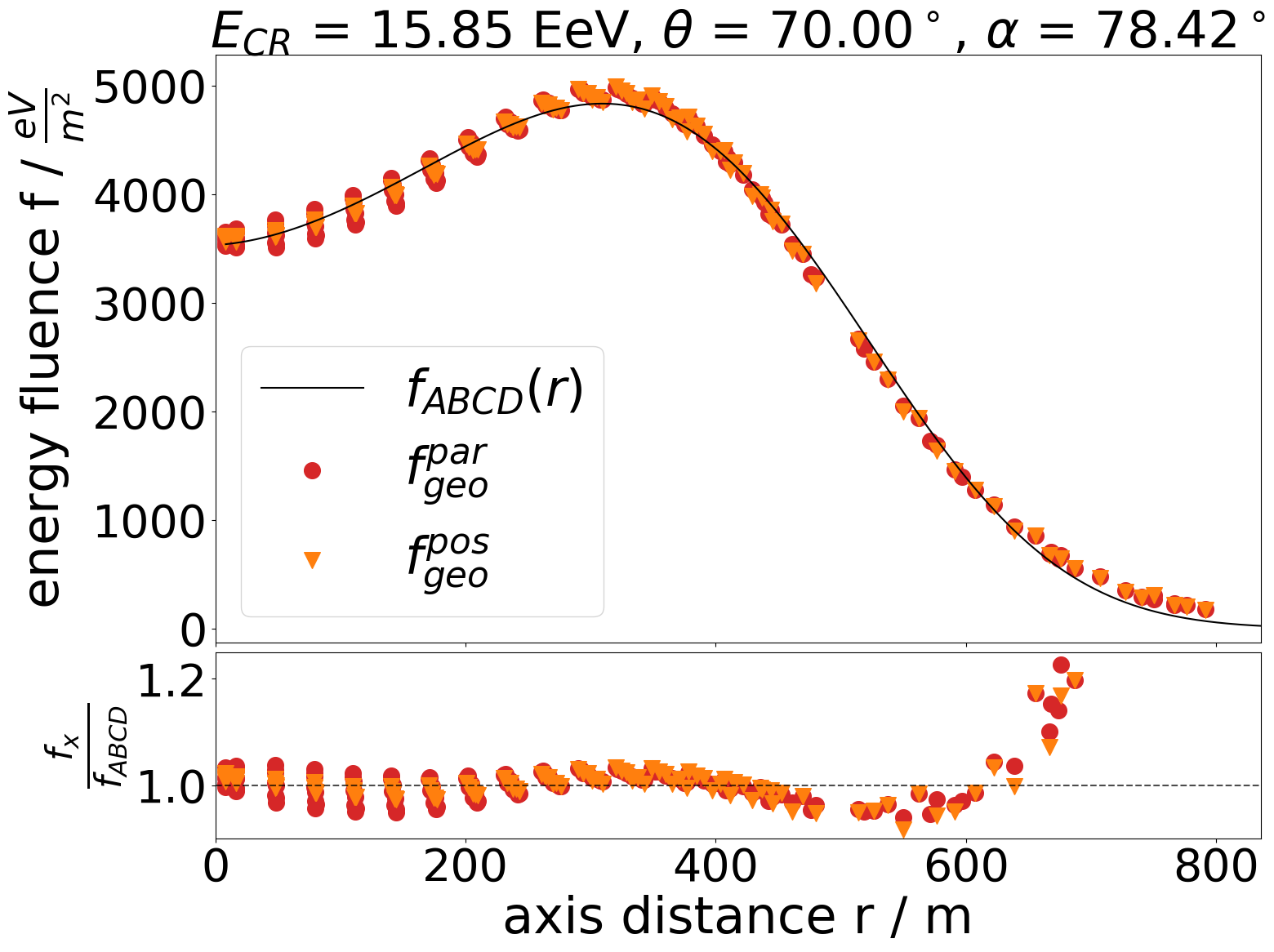}
    \caption{Illustration of a fit of equation (\ref{eqn:ldf}) to the symmetrized signal distribution $f_{\mathrm{geo}}^{\mathrm{par}}$ depicted as red points. Orange points denote geomagnetic energy fluence at each individual simulated position via $f_{\mathrm{geo}}^{\mathrm{pos}}$.}
    \label{fig:ldffit}
\end{figure}
After symmetrization of the signal distribution, we fit a rotationally symmetrical lateral distribution function according to
\begin{equation}
    f_{\mathrm{ABCD}}(r) = A \cdot \exp \left[-B \cdot r - C \cdot r^2 - D \cdot r^3\right] \label{eqn:ldf}
\end{equation}
to the data. The function is unchanged from reference \cite{arena} and is a canonical extension of the exponential of a quadratic function previously used by Tunka-Rex to fit measured amplitudes \cite{TunkaRex}. The achievable fit quality is illustrated in figure \ref{fig:ldffit}, and the red points illustrate the symmetrized signal distribution $f_{\mathrm{geo}}^{\mathrm{par}}$ on the basis of the parameterization of the charge-excess fraction, whereas the orange points illustrate the symmetrized signal $f_{\mathrm{geo}}^{\mathrm{pos}}$ determined at each observer location directly from the signal polarization. The fit function describes the general features of the signal distribution well; nevertheless, we are still investigating alternative functions that might yield an even better fit.

\section{Energy estimation via the geomagnetic radiation energy}
\begin{figure}[t]
    \center
    \includegraphics[width=.49\textwidth]{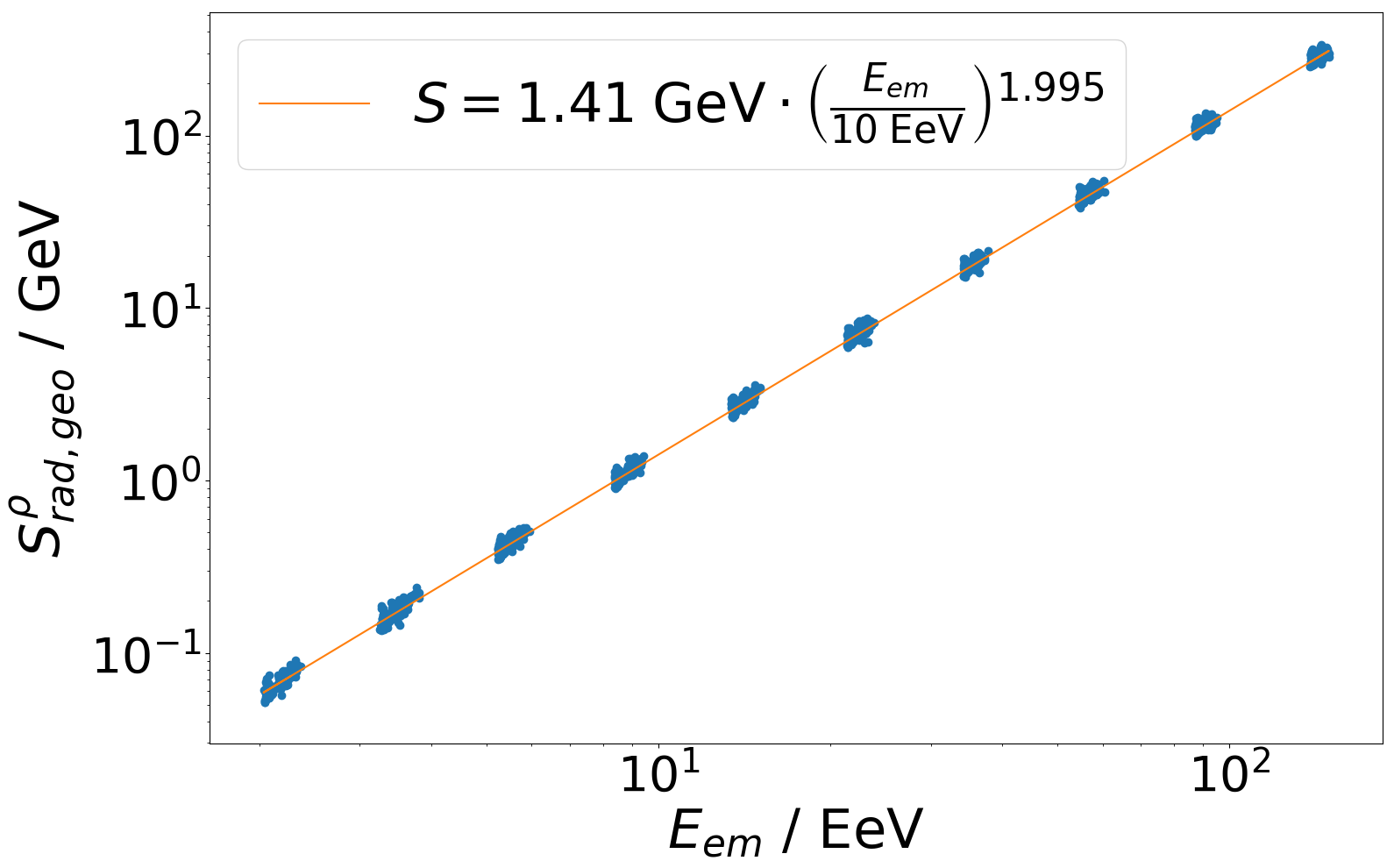}\hfill
    \includegraphics[width=.49\textwidth]{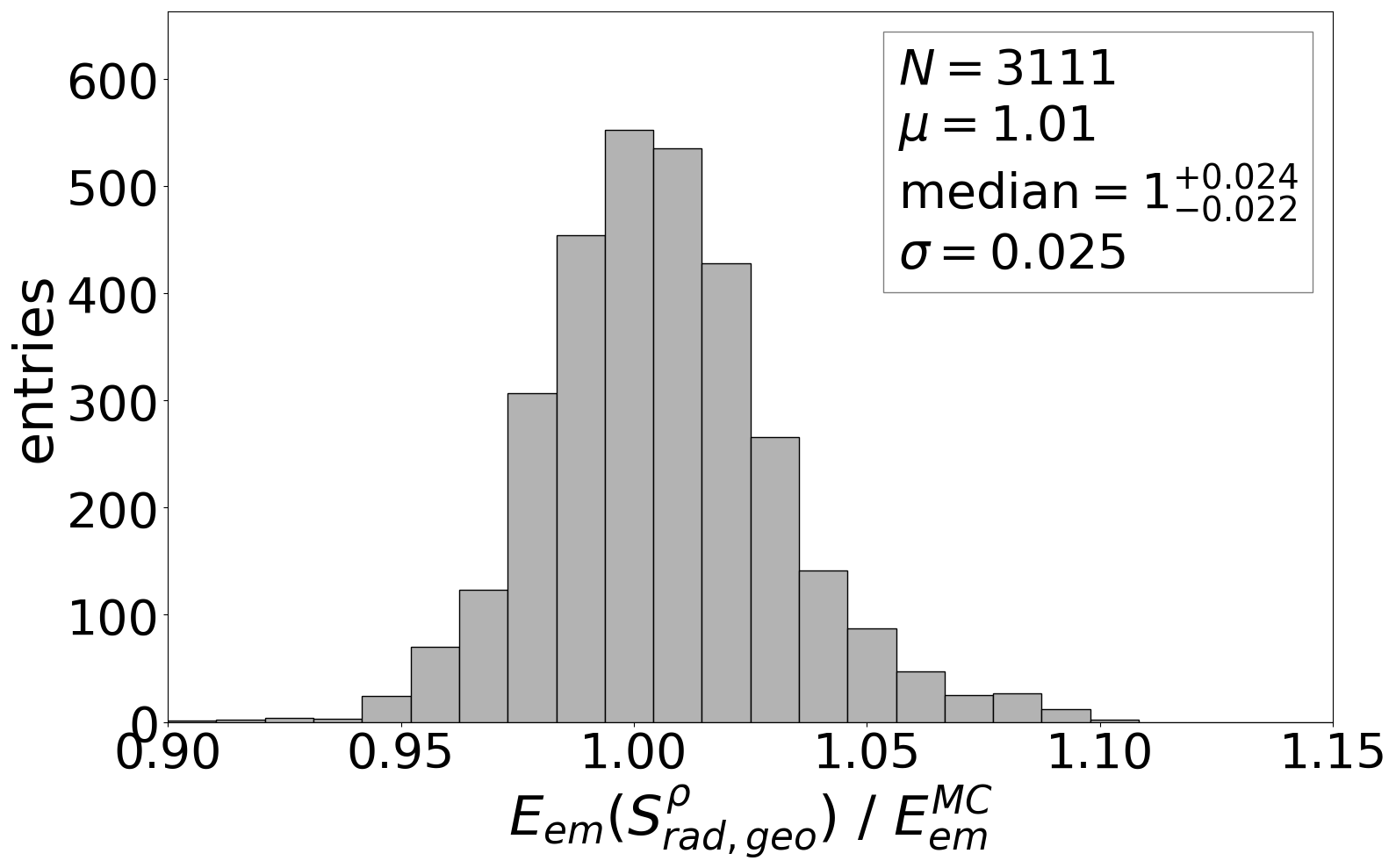}
    \caption{Left: Correlation of the corrected geomagnetic radiation energy as derived with the charge-excess parameterisation and the true electromagnetic energy. Right: Per-event comparison of the electromagnetic energy reconstructed via the corrected geomagnetic radiation energy and true electromagnetic energy.}
    \label{fig:energy_reco}
\end{figure}
In a final step, we determine the radiation energy of the geomagnetic emission component $E_{\mathrm{rad}}^{\mathrm{geo}}$ by numerical integration over the one-dimensional, rotationally symmetric lateral distribution function. The radiation energy needs to be corrected for the scaling of the emission strength with the geomagnetic angle $\alpha$ as well as effects of the atmospheric density, cf.\ reference \cite{GlaserJCAP}:
\begin{equation}
        S_{\mathrm{rad}, \mathrm{geo}}^\rho = \frac{E_{\mathrm{rad}}^{\mathrm{geo}}}{\sin^2(\alpha)} \cdot
        \frac{1}{1 - p_0 + p_0 \cdot \exp\left[p_1 \cdot (\rho - 0.648 \; \mathrm{kg}/\mathrm{m}^3)\right]} \label{eq:corrS}
\end{equation}
This {\em corrected geomagnetic radiation energy} is then an estimator for the energy in the electromagnetic cascade of an extensive air shower $E_{\mathrm{em}}$ according to:
\begin{equation}
S_{\mathrm{rad}, \mathrm{geo}}^\rho = S_{19} \cdot \left(\frac{E_{\mathrm{em}}}{10 \, \text{EeV}}\right)^\gamma \label{eq:Scorrelation}
\end{equation}
The correlation of the true electromagnetic energy with the corrected geomagnetic radiation energy based on the area-integration of $f_{\mathrm{ABCD}}$ fitted on the symmetrized signal distribution $f_{\mathrm{geo}}^{\mathrm{par}}$ is shown in figure \ref{fig:energy_reco} (left) along with the results of a joint fit of equations (\ref{eq:corrS}) and (\ref{eq:Scorrelation}). Simulations for all energies, zenith and azimuth angles and both proton and iron primaries are included in the fit. The fit parameters are denoted in table \ref{tab:fitparams}. The exponent $\gamma$ is in agreement with the expected quadratic scaling for coherent radio emission. The fit predicts a corrected geomagnetic radiation energy of 14.26 MeV at an electromagnetic shower energy of 1 EeV. Figure \ref{fig:energy_reco} (right) illustrates the quality of the reconstruction. The median is essentially unbiased, and the spread is below 3\%. 

\begin{table}[t]
  \caption{Fit parameters of a joint fit of equations (\ref{eq:corrS}) and (\ref{eq:Scorrelation}) to all simulations.}
  \centering\vspace{0.2cm}
  \begin{tabular}{c|cccc}
    \hline
    \hline
    \\[-1em]
    Primary & $S_{19}$ & $\gamma$ & $p_0$ & $p_1$ \\
    \\[-1em]
    \hline
    p, Fe & 1.408 GeV & 1.995 & 0.394 & -2.370 m$^3$/kg \\
    \hline
    \hline
  \end{tabular}
  \label{tab:fitparams}
\end{table}

\section{Conclusions}

We have presented a procedure to symmetrize the signal distribution of radio emission from inclined air showers. This entails a geometrical correction for early-late effects followed by subtraction of the signal contributions due to charge-excess emission to determine the pure geomagnetic emission. We have derived a parameterization of the charge-excess fraction as a function of lateral distance, geometrical distance to shower maximum and atmospheric density at the shower maximum. Alternatively, if signal strength allows, the charge-excess fraction can be determined directly from the data at a given observer position. After symmetrization of the signal distribution, we fit the now rotationally symmetric distribution with a one-dimensional exponential of a cubic polynomial. Integration over this function and correction for well-known geometrical and density effects yields an estimator for the energy of the electromagnetic cascade of an air shower with negligible bias and a spread of less than 3\%. Our approach is fully analytic, documented in closed form, and includes only a small number of parameters. It could be a valuable basis for the development of an event reconstruction for inclined air showers, in particular those to be measured with the Radio Upgrade of the Pierre Auger Observatory \cite{radio_upgrade}.

\end{document}